# Double Kagome bands in a two-dimensional phosphorus carbide $P_2C_3$

Sili Huang, [†] Yuee Xie, [†,*] Chengyong Zhong,[†] Yuanping Chen[†,*]

[†]*School of Physics and Optoelectronics, Xiangtan University, Xiangtan, Hunan, 411105, China*

**Abstract:** The interesting properties of Kagome bands, consisting of Dirac bands and a flat band, have attracted extensive attention. However, the materials with only one Kagome band around the Fermi level cannot possess physical properties of Dirac fermions and strong correlated fermions simultaneously. Here, we propose a new type of band structure --- double Kagome bands, which can realize coexistence of the two kinds of fermions. Moreover, the new band structure is found to exist in a new two-dimensional material, phosphorus carbide $P_2C_3$. The carbide material shows good stability and unusual electronic properties. Strong magnetism appears in the structure by hole doping of the flat band, which results in spin splitting of the Dirac bands. The edge states induced by Dirac and flat bands coexist on the Fermi level, indicating outstanding transport characteristics. In addition, a possible route to experimentally grow $P_2C_3$ on some suitable substrates such as the Ag (111) surface is also discussed.

Kagome band refers to a three-band structure, including two Dirac bands and one flat band[1-5]. The two Dirac bands linearly cross and form a Dirac point at the high-symmetry point $K$, while the flat band quadratically contacts with the Dirac bands at $\Gamma$ point[6-11]. According to energy band theory, both the Dirac bands and flat band are not conventional ones. The former has linear dispersion, and generates massless Dirac fermions with very high mobility[12-18]. On the contrary, the latter generates infinitely massive fermions[19-21]. Their quenched kinetic energies result in Coulomb interactions becoming critical and giving rise to various exotic many-body states[22-28], such as magnetism, superconductivity and Wigner crystallization. Although Kagome band can hold the two types of "paradoxical" bands, the materials with only one Kagome band cannot possess their physical properties simultaneously, because the Dirac point and the flat band cannot appear on the Fermi level at the same time. Therefore, a question is raised: what type of band structure can combine properties of Dirac fermions in the Dirac band and strong correlated fermions in the flat bands?

We propose that double Kagome bands, as shown in Fig. 1(a), can meet the requirement. Around the Fermi level, the flat band in one (green) Kagome band coincidently crosses the Dirac point of the other (purple) Kagome band. In this case, the Dirac fermions and strong correlated fermions coexist at the Fermi level. However, it is very difficult to find double Kagome bands in real materials because of the following reason: a Kagome band is generally produced by two-dimensional (2D) Kagome lattices, but most elements and compounds are not favorable to form Kagome lattice[29-35]. In the lots of 2D materials proposed to date, few of them have Kagome lattice. Even in the few Kagome structures[36-39], to the best of our knowledge, none of them have double Kagome bands.

In this paper, we propose a new two-dimensional (2D) phosphorus carbide $P_2C_3$, as shown in Fig. 2(a), where the gray carbon atoms form Kagome lattice. The new material is selected from a series of 2D structures $X_2C_3$ (X = III, IV and V elements), by using orbital analysis and first-principles calculations. It not only shows good stability, but also exhibits interesting orbital configurations and unique electron properties. The carbon atoms have the same orbitals with those shown in Figs. 1(b) and 1(c), and thus generate double Kagome bands like Fig. 1(a). Because the flat band below the Fermi level is fully occupied, one-hole doping arouses the magnetism of the structure, and leads to spin splitting of the Dirac band in case of no external magnetic field. Coexisting edge states induced by Dirac points and flat band are observed around the Fermi level.

We performed first-principles calculations within the density functional theory (DFT) formalism as implemented in VASP[40]. The electron–electron interactions were treated within a generalized gradient approximation (GGA) in the form of Perdew–Burke–Ernzerhof (PBE) for the exchange–correlation functional[41, 42]. The energy cutoff was set to 600 eV. The atomic positions were fully optimized by the conjugate gradient method[43], the energy and force convergence criteria were set to be $10^{-6}$ eV and $10^{-3}$ eV/Å, respectively. To avoid interaction between adjacent layers, the vacuum distance normal to the layers was kept to 20 Å. Integrations over the Brillouin zone were done with 9×9×1 Monkhorst-Pack k-point grid[44]. To account for the thermal stability, we carried out ab-initio molecular dynamics (AIMD) simulations based on canonical ensemble[45], for which a 4×4 supercell containing 80 atoms was used and the AIMD simulations were performed with a Nose-Hoover thermostat from 300 to 1200 K, respectively.

To get the double Kagome bands in Fig. 1(a), we first consider a tight-binding model for a standard Kagome lattice. When each lattice site has two orthogonal orbitals[46], for example, an in-plane orbital $p_{xy}^1$ and an out-of-plane orbital $p_z$[47, 48], as shown in Figs. 1(b) and 1(c), respectively. Then, the Kagome lattice can be described by the following Hamiltonian:

$$H = \sum_{i,\alpha} \epsilon_{i\alpha} a_{i\alpha}^+ a_{i\alpha} + \sum_{i,j,\alpha} t_{i\alpha,j\alpha} a_{i\alpha}^+ a_{j\alpha}, \qquad (1)$$

where $a_{i\alpha}^+$ and $a_{j\alpha}$ represent the creation and the annihilation operators, respectively. $\epsilon_{i\alpha}$ is the site energy of orbital $\alpha$ at site $i$, and here $\epsilon_1$ and $\epsilon_2$ are used to represent site energies of $p_{xy}^1$ and $p_z$ orbitals, respectively. $t_{i\alpha,j\alpha}$ is the hopping energy between orbitals $\alpha$ at sites $i$ and $j$, and here only the nearest-neighbor hopping parameters are considered, e.g., $t_1$ represents the interactions between $p_{xy}^1$ orbitals, while $t_2$ represents the interactions between $p_z$ orbitals. Because there are no interactions between $p_{xy}^1$ and $p_z$ orbitals, each of them will generate a set of Kagome band. When the parameters in Eq. (1) satisfy $|\epsilon_2 - \epsilon_1| = 2t_2 + t_1$, double Kagome bands appear, as shown in Fig. 1(a), where the green flat band induced by $p_z$ orbital crosses the Dirac point of purple Dirac bands induced by $p_{xy}^1$ orbital. However, in most cases, the orbital parameters in the real materials cannot match the condition, and thus double Kagome bands cannot be found in the 2D single-layer materials to date.

Because most elements and compounds are not favorable to directly form two-dimensional (2D) Kagome lattice, we propose a new lattice to substitute Kagome lattice, as shown in Fig. 2(a), in which the gray atoms form Kagome lattice while the blue atoms linking the gray atoms form hexagonal lattice. Considering the rich bonding chemistry of carbon atom, the gray atoms are fixed as carbon atoms while the blue atoms are selected from III, IV and V elements. Therefore, the primitive cell of the new lattice, labeled as $X_2C_3$, includes three carbon atoms and two X atoms.

We calculate band structures of a series of $X_2C_3$ (X = III, IV and V elements), by using the first-principles methods. The results indicate that this series of structures have double Kagome bands. Especially, for X = Ga, Ge, N, P, the band structures are very close to Fig. 1(a) (see Fig. S2 in the supplementary information (SI)). Then we access the stabilities of these 2D single-layer structures. Their phonon dispersions indicate that only $P_2C_3$ has no soft mode in the phonon spectra (see Figs. 2(c) and S3 in SI). We also examine thermal stability of $P_2C_3$ by performing AIMD simulations in canonical ensemble. After heating up to the targeted temperature 1100 K for 20 ps, we do not observe any structural decomposition (see Fig. 2(d)). The structure reconstruction only occurs at 1200 K during the 20 ps simulation. Therefore, $P_2C_3$ has rather high thermodynamic stability and outstanding dynamic and thermal stabilities. Given that $P_2C_3$ is the only stable structure in $X_2C_3$, we next focus our discussions on this new 2D material. It is a planar single-layer sheet, and its optimized lattice parameters for $P_2C_3$ are $a = b = 5.69$ Å. The bond length between P and C atoms is 1.64 Å. The calculated cohesive energy is 4.60 eV, which intermediate between pure black phosphorus (3.27 eV) and pure graphene (7.67 eV).

The electronic band structure of $P_2C_3$ is shown in Fig. 3(a). It is found that there are double Kagome bands like Fig. 1(a): one is the purple Kagome band in the range [-2.0, 3.0] eV; the other is in the range [-0.1, 8.4] eV, whose flat band is green while Dirac bands are mixture of green and red. The flat band in the top Kagome band is just below the Fermi level, and accidentally cross the Dirac point of the bottom Kagome band. Note that, the 6-fold degenerate crossing point is not protected by any symmetry, and thus a perturbation, for example, an external strain, will shift the flat band away the Dirac point. Additionally, one can find another Dirac bands appear in the range [-4.3, -0.1] eV. A clear examination indicates that there is a small gap (~ 0.1 eV) between this Dirac bands and the green flat band (see the inset in Fig. 3(a)).

To reveal the origination of the band structure, we show partial density of states (PDOS) in Figs. 3(b) and 3(c) for different orbitals of the C and P atoms, respectively. One can find that the double Kagome bands and additional Dirac bands are mainly contributed by $p_{xy}^1$ orbital of C atoms and $p_z$ orbitals of C and P atoms. The colored projected band structure in Fig. 3(a) illustrates detailly the relation between energy bands and the atomic orbitals. The (purple) bottom Kagome band is mainly induced by $p_{xy}^1$ orbitals of C atoms. However, the top Kagome band and the additional Dirac bands are induced by the $p_z$ orbitals of C and P atoms, while the flat band is only related to the $p_z$ orbital of C atoms. Figure 3(d) presents four configurations of atomic orbitals corresponding to four states in the double Kagome bands. It is found that the orbital configurations of the C atoms are similar to those in Figs. 1(b) and 1(c), which are in fact frustrated states induced by orbital frustration[47]. (More details about the projected energy bands of C atoms and configurations of other orbitals can be found in Fig. S4)

To further prove the former orbital analysis, we construct tight-binding model like Eq. (1) to describe $P_2C_3$ and the $p_z$ orbitals of two extra P atoms have been added. The calculated band structure based on the tight-binding model fit the DFT results very well, and the tight-binding parameters for the C atoms are same to those in Fig. 1(a) (see Fig. S5 in SI). These indicate that, the P atoms in $P_2C_3$ not only help the C atoms forming stable Kagome lattice, but also make them obtain standard orbitals in Kagome lattice. More importantly, the $p_{xy}^1$ and $p_z$ orbitals cooperate to get the perfect double Kagome bands.

It is known that the interplay between the band flatness and Coulomb interaction leads to many novel phenomena such as the magnetism[49-51]. In most cases, electrons often remain unpolarized states. However, when electrons partially fill in the flat band, Coulomb interactions are aroused and result in spin polarization[52]. Figure 4 exhibits the band structure of $P_2C_3$ in the case of one-hole doping, i.e., the green flat band in Fig. 3(a) is half filling. In experiment, the doping effect can be achieved by the electrostatic gating[53]. The spin-up and spin-down bands are separated, and the Fermi energy shifts below the upper spin-polarized flat band, such that the lower spin-polarized band is fully occupied. The spin splitting strength is very large (~1.2 eV), roughly on the order of the on-site Coulomb interaction. Interestingly, the spin-down flat band forms Kagome band with the bottom Dirac bands. Our calculations indicate that the total magnetic moment is 0.967 $\mu_B$, in which the magnetic moment of C and P atoms are 1.129 and -0.162 $\mu_B$, respectively. The internal magnetic field leads to spin polarizations of other bands, e.g., the splitting of the Dirac bands around the Fermi level is ~ 0.17 eV. This means that one can obtain spin-polarized Dirac electrons even if no external magnetic field, which is a unique magnetism phenomena of double Kagome bands.

To explore transport properties of $P_2C_3$, we calculate its edge states. Two types of edges are considered: one is zigzag edge on the left side of the nanoribbon in Fig. 5(a), the other is bearded edge on the right side. The edge states for the zigzag edge are shown in Fig. 5(b), while those for the bearded edge are shown in Fig. 5(c). In both cases, there are three edge states around the Fermi level. It is known that, the Berry phases of Dirac point and quadratic contacting point are $\pi$ and $2\pi$[54, 55], respectively. As a result, there is one edge state (labelled as "1") between the Dirac points in the bottom Kagome band, and there are two edge states (labelled as "2") between the quadratic contacting points in the top Kagome band. It is noted that one of the "2" edge states is a flat band and is buried in the bulk flat bands. The coexisting edge states strengthen the electron transport, and thus leads to significant electronic conductivity. Moreover, because of the huge density of state around the Fermi level and the large carrier mobility of the Dirac electrons, the $P_2C_3$ sheet could be a good superconductor with high critical temperature. We expect some further theoretical and experimental studies can be carried out to validate and extend our findings.

Our analyses for structural stabilities already demonstrate the high feasibility to synthesize $P_2C_3$. Considering that many 2D materials have been grown successfully on Au and Ag substrates[56, 57], we explore the possibility to synthesize $P_2C_3$ on Ag (111) substrate. According to our computations, the lattice constants of a 2 × 2 supercell of Ag (111) substrate (a′ = b′ = 5.78 Å) are very close to those of the primitive cell of $P_2C_3$. The mismatch between them is smaller than 1.6%. The adhesion energy between $P_2C_3$ sheet and Ag (111) substrate is -0.164 eV/atom (see Fig. S6 in the SI for more details). It can be comparable with some other 2D materials on the Ag (111) substrate[50]. Therefore, we believe the newly predicted $P_2C_3$ sheet can be synthesized by the pathway similar to those of other 2D materials on the Ag (111) substrate. We also simulate a STM image of $P_2C_3$ sheet on the substrate (see Fig. S7).

In summary, we propose a new band structure, named as double Kagome bands. It includes two sets of Kagome bands, and the flat band in one Kagome band accidently cross the Dirac point in the other Kagome band. In the new type of bands, Dirac fermions and strong correlated fermions can coexist around the Fermi level. A new 2D material $P_2C_3$ possessing double Kagome bands is also proposed, by using first-principles calculations and orbital analysis. Hole doping of the flat band leads to structural strong magnetism, which also induced spin-polarized Dirac electrons without external magnetic field. Two edge-states respectively induced by Dirac points and flat band appear on the Fermi level simultaneously. The $P_2C_3$ sheet shows good stability, and we proposed a promising approach to prepare the new sheet by epitaxial growth on a Ag (111) substrate. To guide the future experimental explorations, we simulated its STM image on the Ag(111) substrate. We hope some experiments can be carried out to validate and extend our findings.

This work was supported by the National Science Foundation of China (Nos. 11474243 and 51376005).

# AUTHOR INFORMATION

**Corresponding Author**

*E-mail: xieyech@xtu.eud.cn; chenyp@xtu.edu.cn.

**ORCID**

Yuanping Chen: 0000-0001-5349-3484

**Notes**

The authors declare no competing financial interest.

# Figure captions

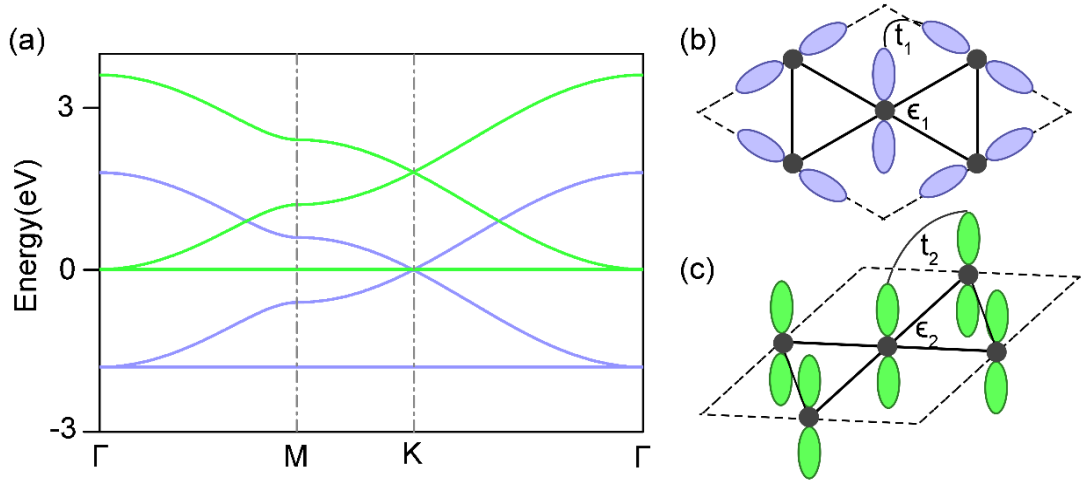

**Figure 1**. (a) Band structure based on tight-binding model in Eq. (1). Schematic views of in-plane $p_{xy}^1$ orbital (b) and out-of-plane $p_z$ orbital (c) in a standard Kagome lattice. $t_1$ and $t_2$ represent the nearest-neighboring hopping energies between in-plane and out-of-plane orbitals, while $\epsilon_1$ and $\epsilon_2$ represent the site energies of the two kinds of orbitals. The purple and green energy bands in (a) are induced by the $p_{xy}^1$, $p_z$ orbitals, respectively, and the parameters are $t_1 = t_2 = 0.6$ eV, $\epsilon_1 = -0.6$ eV, $\epsilon_2 = 1.2$ eV.

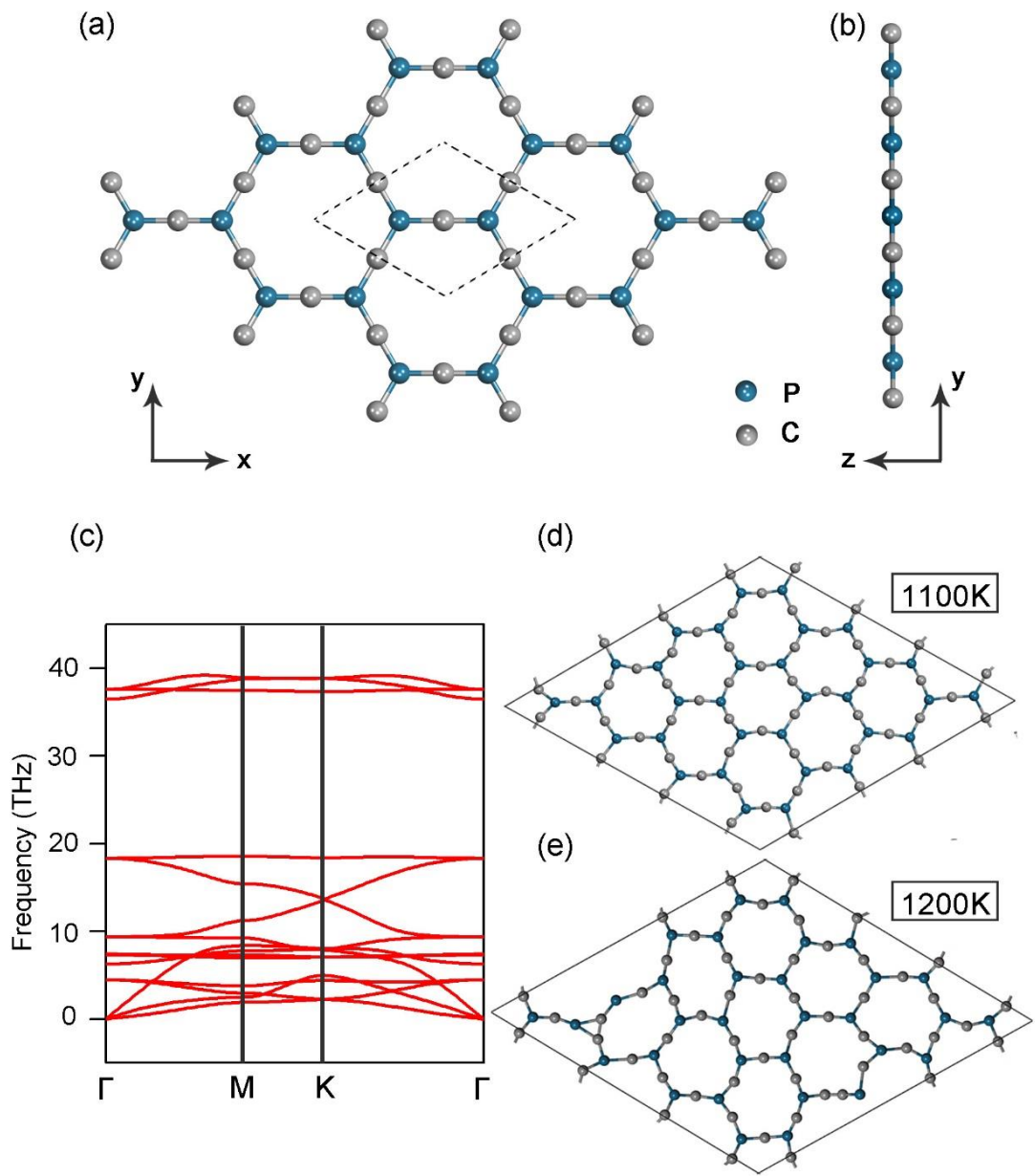

**Figure 2.** Top view (a) and side view (b) of P$_2$C$_3$, where the gray C atoms form Kagome lattice while the blue P atoms form honeycomb lattice. Its primitive cell is shown in the dashed lines in (a). (c) Phonon dispersion of P$_2$C$_3$, where no soft mode is found. Snapshots of the equilibrium structures of P$_2$C$_3$ at the temperatures of (d) 1100 K, (e) 1200 K after 20 ps AIMD simulations.

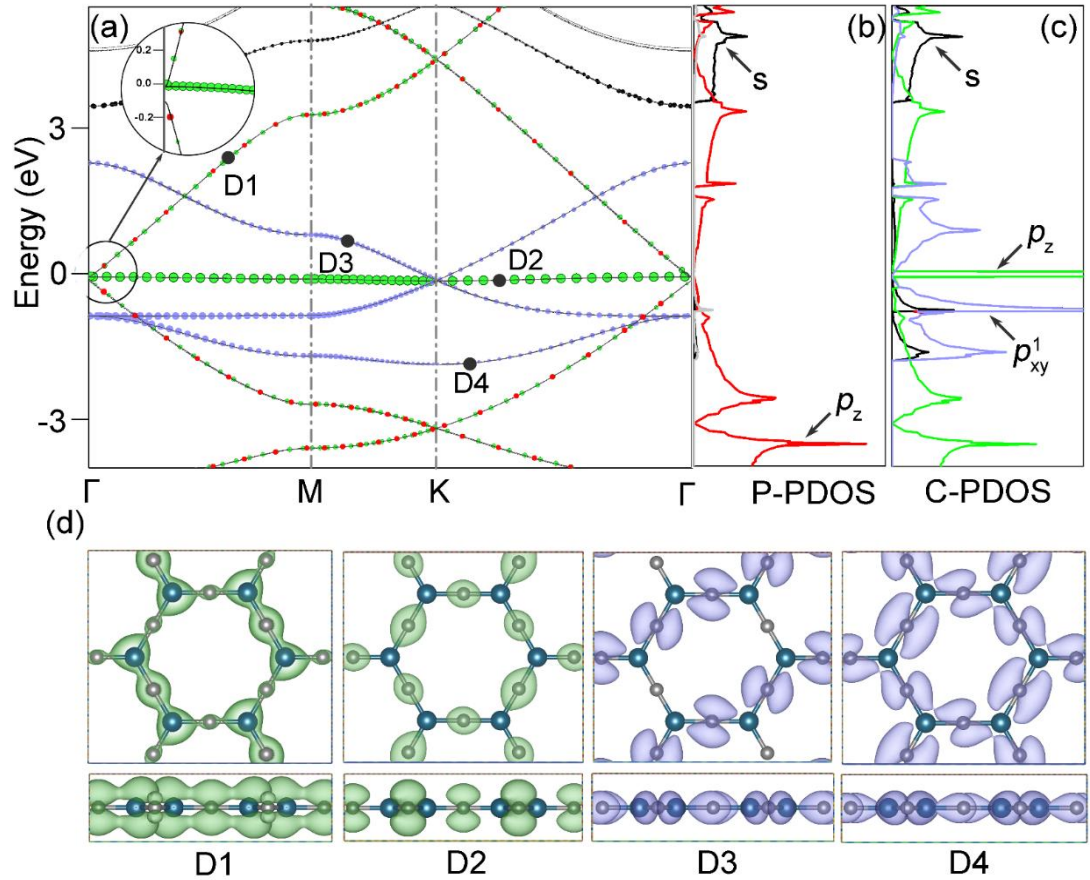

**Figure 3.** (a) The projected band structure of $P_2C_3$ calculated by DFT. PDOS for P atoms (b) and C atoms (c), respectively. The colors of energy bands in (a) correspond to the colors of projected atomic orbitals in (b) and (c). (d) Charge densities for the states from D1 to D4 in (a). The top and bottom panels are top and side views of the states, respectively.

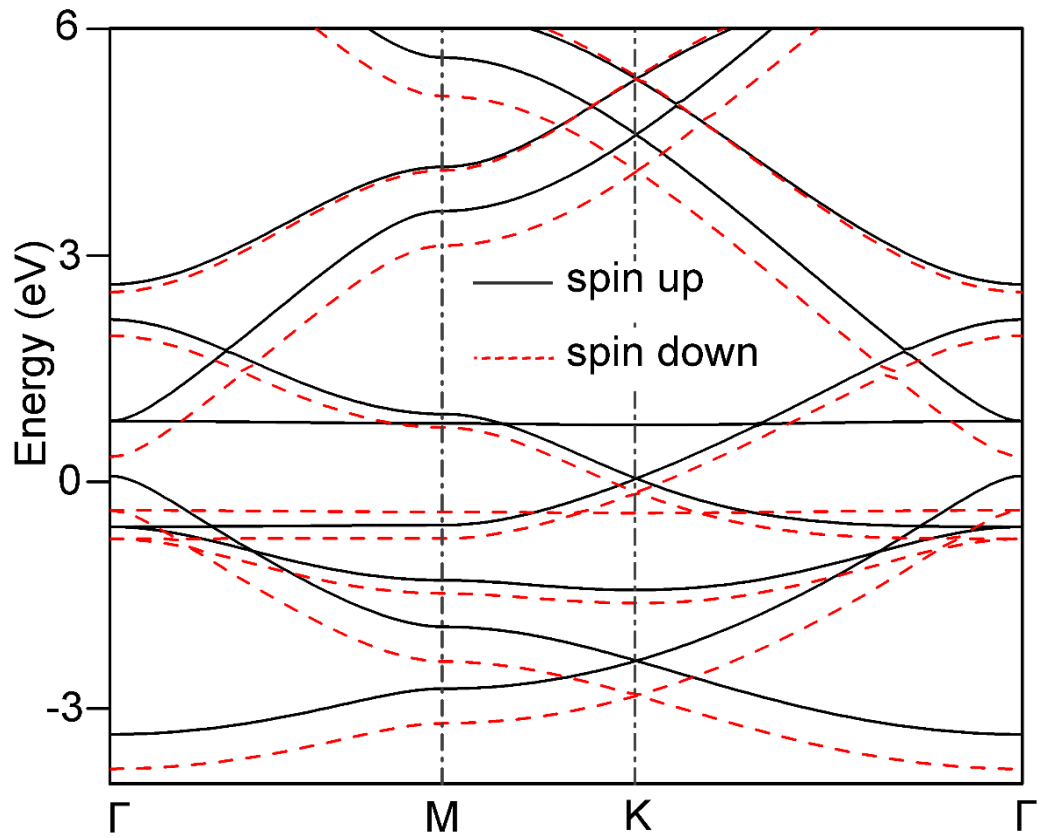

**Figure 4.** The band structure of $P_2C_3$ after one-hole doping, where the solid and dashed lines represent spin-up and spin-down energy bands, respectively. The flat band has a biggest spin splitting about 1.2 eV.

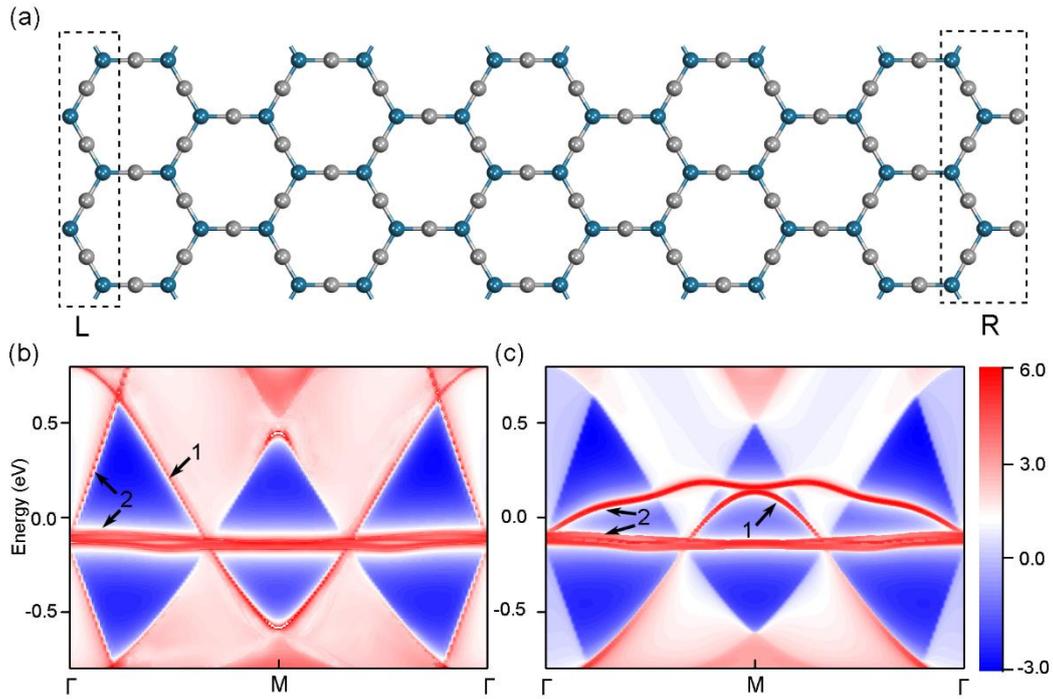

**Figure 5.** (a) Schematic view of a nanoribbon of $P_2C_3$ with two types of edges (zigzag on the left side and bearded on the right side). Edge states for a nanoribbon of $P_2C_3$ with (b) a zigzag edge and (c) a bearded edge, where "1" represents the edge state induced by Dirac point while "2" represents the edge states induced by quadratic contacting point.